\documentclass[twocolumn]{aastex62}
\usepackage{graphics,epsf}
\usepackage{amsmath}                
\usepackage{amsfonts}               
\usepackage{amssymb}                
\usepackage{epsfig}                 
\usepackage{graphicx}
\usepackage{float}
\usepackage{color}
\usepackage[para,online,flushleft]{threeparttable}

\newcommand{\kms}{{~\rm km\; s^{-1}}}

\newcommand{\cm}{{~\rm cm}}
\newcommand{\km}{{~\rm km}}
\newcommand{\s}{{~\rm s}}

\newcommand{\g}{{~\rm g}}

\newcommand{\K}{{~\rm K}}
\newcommand{\erg}{{~\rm erg}}

\newcommand{\days}{{~\rm days}}

\begin{document}

\title{Emission peaks in the light curve of core collapse supernovae by late jets} 

\author{Noa Kaplan}
\affiliation{Department of Physics, Technion, Haifa, 3200003, Israel; noa1kaplan@campus.technion.ac.il; soker@physics.technion.ac.il}

\author[0000-0003-0375-8987]{Noam Soker}
\affiliation{Department of Physics, Technion, Haifa, 3200003, Israel; noa1kaplan@campus.technion.ac.il; soker@physics.technion.ac.il}
\affiliation{Guangdong Technion Israel Institute of Technology, Shantou 515069, Guangdong Province, China}

\begin{abstract}
We build a toy model where the central object, i.e., a newly born neutron star or a black hole, launches jets at late times and show that these jets might account for peaks in the light curve of some peculiar {{{{ (i.e., having unusual light curves) }}}} core collapse supernovae (CCSNe) when the jets interact with the CCSN ejecta. 
We assume that the central object accretes fallback material and launches two short-lived opposite jets weeks to months after the explosion. We model each jet-ejecta interaction as a spherically symmetric `mini explosion' that takes place inside the ejecta. 
{{{{ We assume that each `mini explosion' adds emission that is symmetric in time around the late peak, and with a rise in emission power that has the same slope as that of the main CCSN light curve. In total we use 12 parameters in the toy model.  }}}}
In our toy model late jets form stronger emission peaks than early jets. Late jets with a kinetic energy of only about one percent of the kinetic energy of the CCSN itself might form strong emission peaks. 
We apply our toy model to the brightest peak of the enigmatic CCSN iPTF14hls that has several extra peaks in its light curve. We can fit this emission peak with our toy model when we take the kinetic energy of the jets to be about one to two percent of the CCSN energy, and the shocked ejecta mass to be about three percent of the ejecta mass. 
\end{abstract}


\section{Introduction}
\label{sec:intro}

Observations of polarization in some core collapse supernovae (CCSNe) and morphological features of some supernova remnants, {{{{ e.g., two opposite protrusions termed ` Ears', }}}} strongly suggest that jets play a significant role in many, and possibly in most, CCSNe (e.g., \citealt{Wangetal2001, Maundetal2007, Lopezetal2011, Milisavljevic2013, Gonzalezetal2014, Marguttietal2014, Inserraetal2016, Mauerhanetal2017, GrichenerSoker2017, Bearetal2017, Garciaetal2017,  LopezFesen2018}). 
\cite{Boseetal2019}, as a recent example, study the Type II-P CCSN ASASSN-16at (SN 2016X) and argue that the nebular-phase Balmer emission suggests that the $^{56}$Ni in this CCSN has a bipolar morphology. Such a morphology is expected in jet-driven explosions (e.g., \citealt{Orlandoetal2016, BearSoker2018}).

From the theoretical side, the problems of the delayed neutrino explosion mechanism (e.g., \citealt{Papishetal2015, Kushnir2015b}) brought the suggestion that the jittering jets explosion mechanism explodes most, or even  all, CCSNe (e.g. \citealt{PapishSoker2011, GilkisSoker2015}), including super-energetic (or super-luminous) CCSNe (\citealt{Gilkisetal2016, Soker2017RAA}; for a review see \citealt{Soker2016Rev}).
There are many studies of jets in  CCSNe (e.g., \citealt{Khokhlovetal1999, Aloyetal2000, Hoflich2001, MacFadyen2001, Obergaulingeretal2006, Burrows2007, Nagakuraetal2011, TakiwakiKotake2011, Lazzati2012, Maedaetal2012, LopezCamaraetal2013, Bromberg_jet, Mostaetal2014, LopezCamaraetal2014, Itoetal2015, BrombergTchekhovskoy2016, LopezCamaraetal2016, Nishimuraetal2017, Fengetal2018, Gilkis2018}), but in most cases these studies consider jets to play a role only in rare types of CCSNe where the pre-collapse core is rapidly rotating.
{{{{ Namely, only in rare cases when the collapsing core forms a long-lived accretion disk around the newly born NS. }}}} The jittering jets explosion mechanism that works via a negative feedback, works for both slowly and rapidly rotating pre-collapse cores. The neutrino heating does play a significant role in the jittering jets explosion mechanism, at least in regular (not super-energetic) CCSNe \citep{Soker2018KeyRoleB, Soker2019SASI}. 

One outcome of a rapidly rotating pre-collapse core might be a late {{{{ (weeks to months post-explosion) }}}} fallback of gas that forms an accretion disk around the newly born neutron star (NS) or black hole, and the accretion disk launches late jets. 
In the present study we use a non-spherical toy model to examine the possibility that intermittent late jets energize peaks in the light curve of some rare types of CCSNe. 
In Section \ref{sec:ToyModel} we present the properties of the CCSN and its ejecta, and our approach that considers each jet-ejecta interaction as `mini-explosion'. In Section \ref{sec:Peaks} we present the extra emission of the peaks that the mini-explosions energize. In Section \ref{sec:iPTF14hlsToy} we apply our results to the enigmatic SN~iPTF14hls and discuss the scenarios that allow for late intermittent jets. 
{{{{ The extraordinary type II SN~iPTF14hls has some unusual properties, like a very slow evolution of the light curve with a rapid decay only after over two years, a large total emitted energy, and several peaks in its light curve \citep{Arcavietal2017Nature, Sollermanetal2019}. We try to explain one of the peaks (the other peaks are consistent with the model of \citealt{Wangetal2018}). }}}}
We summarize our results in Section \ref{sec:summary}.

\section{The toy model}
\label{sec:ToyModel}

\subsection{Bare SN light curve}
\label{sec:SN}

In this section we describe the {{{{ artificial SN light curve that we build, based in part }}}} on the literature, and that we scale with expressions that include photon diffusion and recombination.
We assume that the SN explosion is spherically symmetric and that cooling is due to photon diffusion and adiabatic expansion, while hydrogen recombination releases energy. 

\begin{figure}[ht!]
	\centering
	\includegraphics[trim=3.5cm 8cm 4cm 8cm ,clip, scale=0.6]{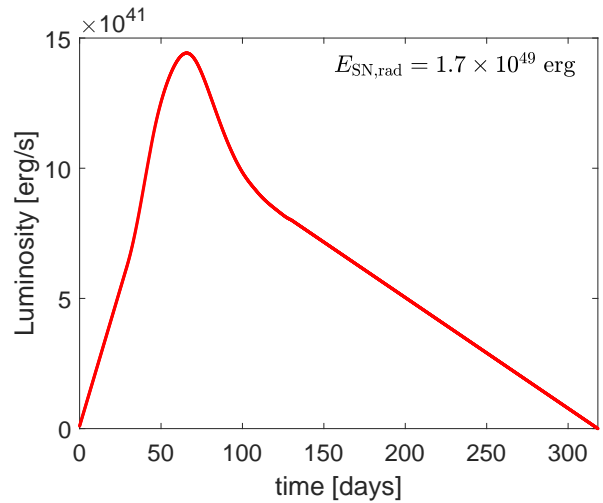}
	\caption{The light curve of the CCSN SN itself in our toy model. The shape {{{{ of the peak of the CCSN light curve }}}} is based on The Open Supernova Catalog \citep{Guillochonetal2017}, while the time of maximum luminosity and the maximum luminosity are scaled by equations (\ref{eq:tSN}) and (\ref{eq:LSN}), respectively. The total energy that the bare SN radiates in our toy model is $E_{\rm SN} = 1.7 \times 10^{49} \erg$.
	}
	\label{fig:SNLightCurve}
\end{figure}

{{{{ We present the SN light curve that we use for our toy model in Fig. \ref{fig:SNLightCurve}. In constructing this light curve we used the photometric data of SN 2008ax from The Open Supernova Catalog (\citealt{Guillochonetal2017}) to fit the peak. We chose this CCSN because it is one of many CCSNe that have clear peak in their light curve, but we could have chosen many other light curves. We then built an artificial continuation to early and late times. We assume that this light curve includes all contributions beside that of the jets that we add later, namely, it includes the initial heat of the ejecta, recombination of the ejecta, and radioactivity. Our results are not sensitive to the exact shape of the CCSN light curve; it serves us as a background light curve. }}}}
   
As well, instead of the luminosity and time of SN 2008ax we scale the timescale $t_{\rm SN}$ and typical luminosity $L_{\rm SN}$ of this bare SN light curve in our toy model by using  equations (8) from \cite{Kasen09}. More specifically, here we take $t_{\rm SN}$ to be the time from explosion to the peak luminosity, and we take $L_{\rm SN}$ to be the peak luminosity. 
The scaled timescale and luminosity of the bare SN light curve we use in the toy model read then
\small
\begin{eqnarray}
\begin{aligned} 
& t_{\rm SN} = 66 
\left(\frac{E_{\rm SN}}{2\times 10^{51} \erg}\right)^{-1/6} \left(\frac{M_{\rm SN}}{10 M_{\odot}}\right)^{1/2}
\\ & \times 
\left(\frac{R_0}{300 R_{\odot}}\right)^{1/6} \left(\frac{\kappa}{0.38  \cm^2 \g^{-1}}\right)^{1/6} \left(\frac{T_{\rm ph}}{6000 \K}\right)^{-2/3} {~\rm d} ,
\end{aligned}
\label{eq:tSN}
\end{eqnarray}
\normalsize
and 
\small
\begin{eqnarray}
\begin{aligned} 
& L_{\rm SN} = 1.4\times 10^{42} 
\left(\frac{E_{\rm SN}}{2\times 10^{51} \erg}\right)^{5/6} \left(\frac{M_{\rm SN}}{10 M_{\odot}}\right)^{-1/2}
\\ & \times
\left(\frac{R_0}{300 R_{\odot}}\right)^{2/3} \left(\frac{\kappa}{0.38 \cm^2 \g^{-1}}\right)^{-1/3} \left(\frac{T_{\rm ph}}{6000 \K}\right)^{4/3} \erg \s^{-1},
\end{aligned}
\label{eq:LSN}
\end{eqnarray}
\normalsize
where $E_{\rm SN}$ is the explosion energy, $M_{\rm SN}$ is the ejecta mass, $R_0$ is the initial radius of the SN progenitor star (at $t=0$), 
$\kappa$ is the opacity of the ejecta (scaled according to, e.g., \citealt{NagyVink2016}), and $T_{\rm ph}$ is the temperature of the ejecta photosphere. 
We take $T_{\rm ph} = 6000 \K$ since below this temperature the outer layer of the SN ejecta cools enough to allow recombination of the hydrogen, resulting in the ejecta becoming transparent above the recombination front \citep{Kasen09}. 
From these parameters we also calculate the typical ejecta velocity $v_{\rm SN} \approx (2E_{\rm SN}/M_{\rm SN})^{1/2} = 4500 \km \s^{-1}$. 

\subsection{The mini-explosions toy model}
\label{sec:jet}

We present a toy model for light curves of CCSNe that suffer extra power from jets that the central object launches at late times. We assume that each jet causes a `mini-explosion' in the region where it interacts with the SN ejecta. 
{{{{ We build a general toy model, with an artificial SN background light curve (Fig. \ref{fig:SNLightCurve}), that can in principle be applied to most types of CCSNe. For each specific case one will have to use the appropriate SN background light curve before adding the mini-explosions of the jets. }}}}

To make the toy model as simple as possible, e.g., a minimum number of parameters,  we build the shape of the light curve of each mini-explosion as follows. We assume that the rise of the mini-explosion to maximum luminosity has the same shape as the rise of the bare SN light curve to maximum (Fig. \ref{fig:SNLightCurve}).
We then take the decline of the mini-explosion from maximum to be symmetric to its rise (rather than having a longer tail as the bare SN light curve has). Below we describe our scaling of the timescale and luminosity of the mini explosion. {{{{ In Table \ref{Table:Parameters} we list the parameters that we use in the toy model, as well as the assumptions that we use to define these parameters and in using some equations. }}}}  
\begin{table*} 
\caption{The different independent and calculated variables in the toy model.  }
\begin{tabular}{lccccc}
\hline
Parameter                         & Symbols                  &1$^{\rm st}$ appearance               & Value                                                         & Assumptions\\
\hline
SN explosion energy               & $E_{\rm SN}$             & eq. (\ref{eq:tSN}), (\ref{eq:LSN})   & $2 \times 10^{51} \erg$                                       & a,c \\
SN ejecta mass                    & $M_{\rm SN}$             & eq. (\ref{eq:tSN}), (\ref{eq:LSN})   & $10 M_\odot$                                                  & a,c \\  
Progenitor radius                 & $R_0$                    & eq. (\ref{eq:tSN}), (\ref{eq:LSN})   & $300 R_\odot$                                                 & a\\ 
Photosphere temperature           & $T_{\rm ph}$             & eq. (\ref{eq:tSN}), (\ref{eq:LSN})   & $6000 \K$                                                     & a,f,g\\
Ejecta Opacity                    & $\kappa$                 & eq. (\ref{eq:tSN}), (\ref{eq:LSN})   & $0.38 \cm^2 \g^{-1}$                                          & b,f\\
Radial location of mini-explosion & $\beta$                  & eq. (\ref{eq:Rtj0})                  & $1/2$                                                         & h\\
Time of mini-explosion            & $t_{\rm j,0}$            & eq. (\ref{eq:Rtj0})                  & $87$~day, $29$~day                                        & h\\
Cocoon Opacity                    & $\kappa_c$               & eq. (\ref{eq:jet})                   & $0.38 \cm^2 \g^{-1}$                                          & b,f\\
half-opening angle of jet         & $\alpha_{\rm j}$         & Above eq. (\ref{eq:Rc})              & $5^{\circ}$                                                   & i \\
half-opening angle of cocoon      & $\alpha_{\rm c}$         & eq. (\ref{eq:jet2})                  & $30^{\circ}$                                                  & i \\
Cocoon to SN energy ratio         & $\epsilon_E$             & eq. (\ref{eq:jet2})                  & $0.003$                                                       & i \\
Jet terminal velocity             & $v_{\rm L}$              & eq. (\ref{eq:t_L})                   & $10^{10} \cm \s^{-1}$                                         & i \\
\hline 
Calculated variables              &                                                                 &                                      &                                &  \\ 
\hline
Velocity of SN ejecta             & $v_{\rm SN}=\left(\frac{2E_{\rm SN}}{M_{\rm SN}}\right)^{1/2}$  & Below eq. (\ref{eq:LSN})             & $4500 \kms$                    & c\\
SN Radius of the interaction      & $R_{\rm s}(t_{\rm j,0})=\beta R(t_{\rm j,0})$                   & eq. (\ref{eq:Rtj0})                  & $24000 R_\odot, 8200 R_\odot$  & h \\
Cocoon initial radius             & $R_{\rm c,0}=R_{\rm s}(t_{\rm j,0}) \sin {\alpha_{\rm j}}$      & eq. (\ref{eq:jet})                   & $2100 R_\odot, 710 R_\odot$    & d \\
Cocoon energy                     & $E_{\rm c}=\epsilon_E E_{\rm SN}$                               & eq. (\ref{eq:jet}), (\ref{eq:jet2})  & $6\times10^{48}\erg$           & d,e\\
Cocoon mass                       & $M_{\rm c}=\epsilon_V  M_{\rm SN}$                              & eq. (\ref{eq:jet}), (\ref{eq:jet2})  & $0.67 M_\odot$                 & d,e\\
One cocoon to SN mass ratio       & $\epsilon_V=0.5\left(1-\cos{\alpha_{\rm c}}\right)$             & eq. (\ref{eq:jet2})                  & $0.067$                        & i\\
Jet launch time                   & $t_{{\rm L}} = t_{\rm j,0} - R_{\rm s}(t_{\rm j,0})/v_{\rm L}$  & eq. (\ref{eq:t_L})                   & $85 \days, 28 \days$           & i\\
\hline
\end{tabular}
\newline
{{{{ Notes:  The upper part of the table lists the independent parameters, and the lower part lists the calculated variables, both for the models we present in section \ref{sec:Peaks}. The relevant main assumptions that we mark in the fifth column are as follows. (a) Spherical CCSN explosion;  (b) Constant opacity; (c) Homologous expansion of SN; (d) Spherical mini-explosions; (e) Homologous expansion of each cocoon; (f) Optically thick medium; (g) Recombination determines the photosphere radius of the SN and $T_{\rm ph}$ is fixed near the ionization temperature; (h)  The jets deposit their energy in a small volume and in a short time inside the ejecta; (i) The central object launches narrow jets at about its escape velocity, having a mass of about 10-20 percent of fall back mass. }}}}
\label{Table:Parameters}
\end{table*}

Although each jet propagates along a specific direction and is most likely to inflate an elongated `cocoon', in our toy model we assume that each mini-explosion is spherically symmetric around the point where the jet interacts with the ejecta. We also assume that cooling is due to photon diffusion and adiabatic expansion, much as the SN explosions themselves are. These assumptions allow us to use results for a SN explosion to determine the luminosity and timescales of each jet's mini-explosion. 

In each jet-launching episode the central engine launches two opposite jets. We here present the relevant quantities for one jet. We assume that the jet deposits its energy via a shock wave inside the SN envelope at a radius that {{{{ is a fraction $\beta$ of the SN radius }}}}  
\begin{equation}
R_{\rm s}(t_{\rm j,0}) = \beta R(t_{\rm j,0})= \beta \left(v_{\rm SN} t_{\rm j,0} +R_0 \right), 
\label{eq:Rtj0}
\end{equation}
where $t_{\rm j,0}$ is the time the jet interacts with the ejecta, $R(t_{\rm j,0})$ is the radius of the SN at the time of interaction, $v_{\rm SN}$ is the velocity of the ejecta, and $R_0$ is the initial radius of the SN itself at its explosion time (see Section \ref{sec:SN}). The radius of the SN is $R(t)= v_{\rm SN} t +R_0$.
We take $\beta = 1/2$, i.e., in our toy model the mini-explosion takes place at half the radius of the SN ejecta outer edge.

\begin{figure}[t]
	\centering
	\includegraphics[trim=31cm 12cm 31cm 5cm ,clip, scale=0.13]{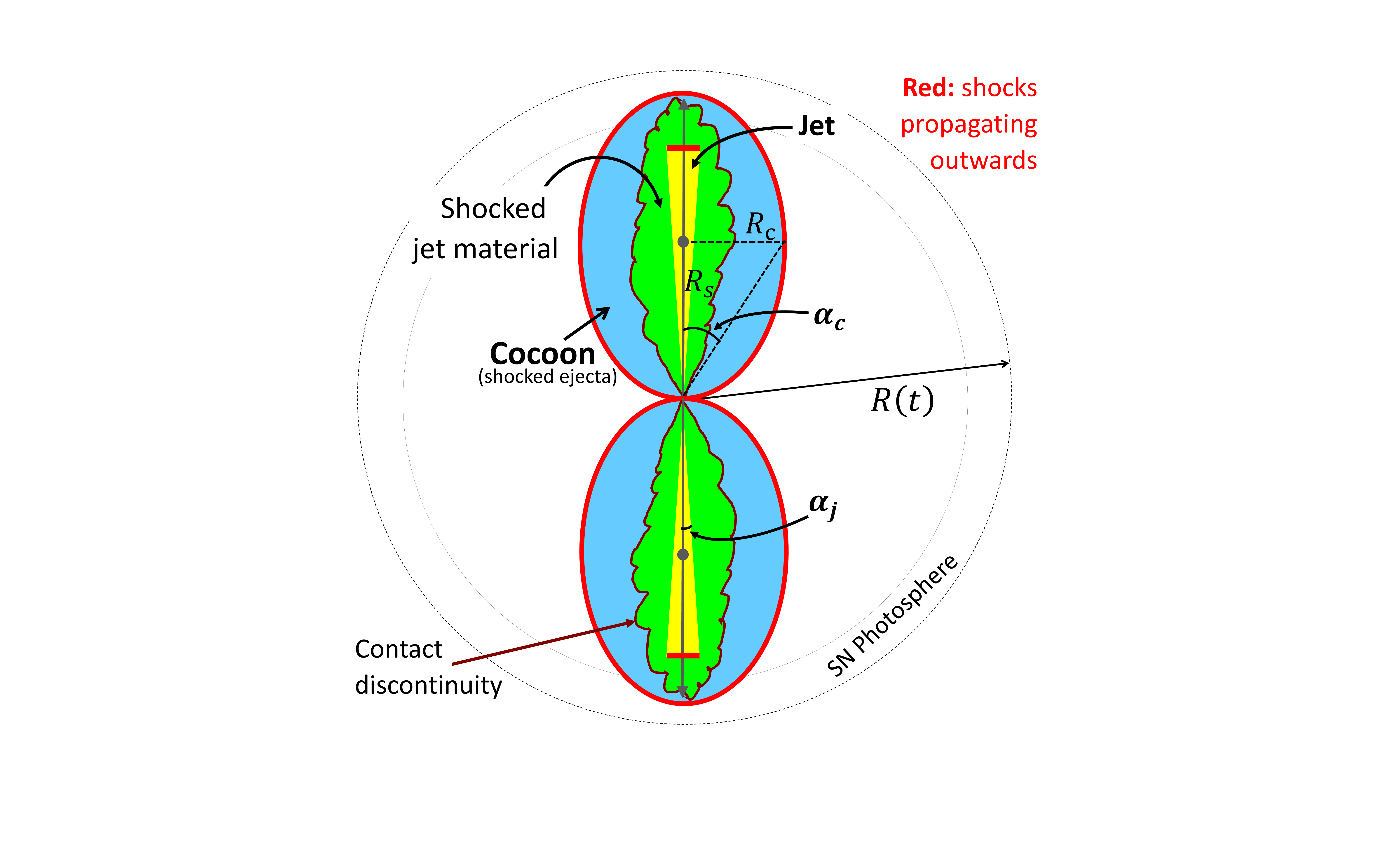}
	\caption{A Schematic illustration (not to scale) of the `mini-explosions' of two opposite jets. }
	\label{fig:schecmatic}
\end{figure}

Each jet shocks the ejecta to form a hot shocked ejecta zone, the so called `cocoon'. The shocked jets' gas and the cocoon, with a contact discontinuity between them, expand to shock more ejecta gas. The hot regions cool by adiabatic expansion and radiation that adds to the total SN light curve. In Fig. \ref{fig:schecmatic} we schematically present the interaction of the jets with the ejecta.

In our toy model of the mini-explosion we consider only photon diffusion through the cocoon. Since the timescale of a jet's mini-explosion is shorter than the SN explosion timescale and occurs under different conditions, e.g., deeper in the hotter envelope {{{{ where the temperature is $ T> 10^4 \K$, }}}} we assume that {{{{ most of }}}} the cocoon does not reach the recombination phase during the relevant time of the mini-explosion. So we neglect the recombination process in the mini-explosion. In any case, any contribution of recombination is included already in the light curve of the SN itself. We therefore use equations (4) from \cite{Kasen09} for the mini-explosions, rather than their equations (8) that we use for the SN explosion itself (Section \ref{sec:SN})
\begin{eqnarray}
\begin{aligned} 
& t_{\rm c} = \left( \frac{3}{2^{5/2} \pi^2 c} \right)^{1/2} E_{\rm c}^{-1/4} M_{\rm c}^{3/4} \kappa_{\rm c}^{1/2},\\ 
& L_{\rm c} = \frac{2 \pi c}{3}
E_{\rm c} M_{\rm c}^{-1} R_{{\rm c},0} \kappa_{\rm c}^{-1},
\end{aligned}
\label{eq:jet}
\end{eqnarray}
where $E_{\rm c}$, $M_{\rm c}$, $R_{{\rm c},0}$ and $\kappa_{\rm c}$, are the energy that the jet deposits into the cocoon, the mass of the shocked cocoon, the initial radius of the shocked cocoon, and the opacity of the cocoon, respectively. 

We assume that shortly after the photons diffuse out from the cocoon they escape from the SN photosphere, and so we do not add the diffusion time of photons through the ejecta outside the cocoon.

For the initial radius of the cocoon we take the radius of the jet's cross section at the place where the jet interacts with the ejecta, namely,
$R_{{\rm c},0} = R_{\rm s}(t_{\rm j,0}) \sin {\alpha_{\rm j}}$, where $\alpha_{\rm j}$ is the half opening angle of the jet. Taking $R_{\rm s}(t_{\rm j,0})$ from equation (\ref{eq:Rtj0}) we can write
\begin{equation}
R_{{\rm c},0} = \beta \left[\left(\frac{2E_{\rm SN}}{M_{\rm SN}}\right)^{1/2}t_{\rm j,0}+R_0 \right]  \sin{\alpha_{\rm j}},
\label{eq:Rc}
\end{equation} 
where $t_{\rm j,0}$ is the time the mini-explosion occurs.

We express the parameters of the cocoon relative to the parameters of the SN itself 
\begin{eqnarray}
\begin{aligned} 
& E_{\rm c} = \epsilon_E E_{\rm SN} \\
& M_{\rm c} = \epsilon_V M_{\rm SN} =0.5 \left(1-\cos{\alpha_{\rm c}}\right)  M_{\rm SN} ,
\end{aligned}
\label{eq:jet2}
\end{eqnarray}
where $\epsilon_E$ is the ratio between the energy that the jet deposits into one cocoon and the total SN energy, $\epsilon_V =0.5 \left(1-\cos{\alpha_{\rm c}} \right)$ is the fraction of the ejecta mass that ends up in one cocoon. This expression and Fig. \ref{fig:schecmatic} define $\alpha_{\rm c}$.  

We substitute equations (\ref{eq:Rc}) and (\ref{eq:jet2}) into equations (\ref{eq:jet}) and scale quantities to obtain the timescale and the luminosity of a mini-explosion   
\small
\begin{eqnarray}
\begin{aligned} 
& t_{\rm c} = 56 
\left(\frac{\epsilon_V}{0.067}\right)^{3/4}
\left(\frac{\epsilon_E}{0.01}\right)^{-1/4}
\left(\frac{E_{\rm SN}}{2\times 10^{51} \erg}\right)^{-1/4} \\ & \times
\left(\frac{M_{\rm SN}}{10 M_{\odot}}\right)^{3/4}
\left(\frac{\kappa_{\rm c}}{0.38 \cm^2\g^{-1}}\right)^{1/2} {~\rm d}, 
\end{aligned}
\label{eq:tjet}
\end{eqnarray}
\normalsize
and
\small
\begin{eqnarray}
\begin{aligned} 
& L_{\rm c} = 4.2\times 10^{41} 
\left(\frac{\epsilon_V}{0.067}\right)^{-1}
\left(\frac{\epsilon_E}{0.01}\right) \left(\frac{\sin{\alpha_{\rm j}}}{0.087}\right) \\ & \times
\left(\frac{\beta}{0.5}\right)
\left(\frac{M_{\rm SN}}{10 M_{\odot}}\right)^{-3/2}
\left(\frac{E_{\rm SN}}{2\times 10^{51} \erg}\right)^{3/2}
 \\ & \times
\left(\frac{\kappa_{\rm c}}{0.38  \cm^2\g^{-1}}\right)^{-1}
\left(\frac{t_{\rm j,0}}{100 {~\rm d}}\right) \erg \s^{-1}, 
\end{aligned}
\label{eq:Ljet}
\end{eqnarray}
\normalsize
respectively, where the opacity of the cocoon is $\kappa_{\rm c} = 0.38 \cm^2\g^{-1}$, as we assumed for the SN ejecta, and we neglect $R_0$ in using equation (\ref{eq:Rc}) as we study late times when the SN radius is much larger than its initial radius $R_0$.

The normalization in equation (\ref{eq:Ljet}) is for $\alpha_{\rm j} = 5^{\circ}$, and for $\alpha_{\rm c} = 30^{\circ}$ that we take from the following consideration. 
We assume that the half-width of the two opposite protrusions in some SN remnants that \cite{GrichenerSoker2017} assume to result from jets, represent the angle $\alpha_{\rm c}$. From their results we scale with $\alpha_{\rm c} = 30^{\circ}$.  

{{{{ Considering the assumption of a spherical cocoon. If the bubble and cocoon are not spherical, then there are two opposite effects. Photons can escape faster along the shorter dimension where the optical depth is lower. On the other hand, the cocoon can expand faster  along the shorter dimension (relative to its initial dimension), therefore, suffering more adiabatic cooling. As we do not expect the cocoon to be very elongated, we take these two opposite effects to imply that our assumption of a spherical cocoon is adequate to the present study.  }}}}

\section{The peaks of the jets' mini-explosions}
\label{sec:Peaks}

We present results of the toy model for one set of parameters.
In section \ref{sec:iPTF14hlsToy} we show that the toy model might crudely fit the third peak (as defined in \citealt{Wangetal2018}) in the light curve of iPTF14hls. First we recall that our derivation of equation (\ref{eq:Ljet}) is for one jet, but in each jet-launching episode there are two jets. In calculating the total emission resulting from one launching episode we will take the luminosity to be $L_{\rm 2c}=2 L_c$. 

Our set of parameters for these particular demonstrative cases are as follows. 
For the bare supernova properties we take the light curve as given in Fig. \ref{fig:SNLightCurve} scaled with a total kinetic energy of 
$E_{\rm SN}= 2 \times 10^{51} \erg$ and an ejecta mass of $M_{\rm SN} = 10 M_{\odot}$. 

For the jets' half opening angle we take $\alpha_j=5^\circ$ {{{{ (see definition in Fig. \ref{fig:schecmatic} and values of parameters in Table \ref{Table:Parameters}).  }}}}

\begin{figure}[t!]
	\centering
	\includegraphics[trim=3.5cm 8cm 4cm 8cm ,clip, scale=0.6]{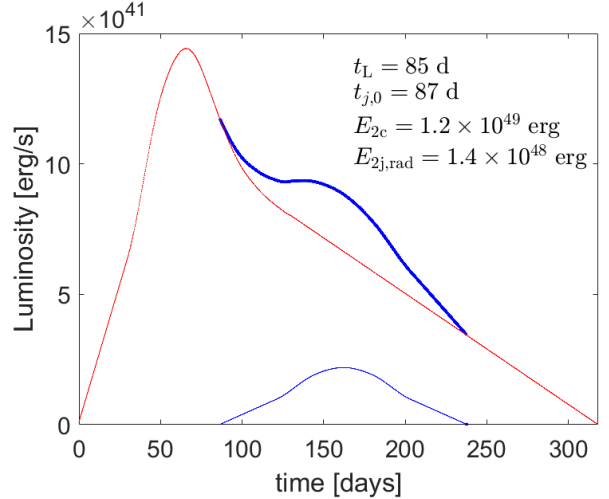}
	\caption{The light curve resulting from a late jet-ejecta interaction in our toy model. The thin-red line is the bare SN light curve from Fig. \ref{fig:SNLightCurve}, while the thin-blue line (lower line) is the contribution of the jet-ejecta interaction of the two opposite jets combined. The thick-blue line is the combined light curve during the activity phase of the jets. 
		$t_{{\rm L}}$, $t_{\rm j,0}$, $E_{2{\rm c}}$, and $E_{\rm {2j,rad}}$ are 
		the launching time of the two opposite jets, the 'mini-explosion' time of the jets, the energy that the two jets deposit to the cocoons, and the extra energy radiated by the two cocoons. 
		For the other parameters of this case see text and Table \ref{Table:Parameters}. 
	}
	\label{fig:SN&jet}
\end{figure}

The first case we present has $\epsilon_E=0.003$, i.e., one jet deposits an energy of 
$E_{\rm c} = 6 \times 10^{48}  \erg$ into its cocoon (see equation \ref{eq:jet2}), and $\epsilon_V = 0.067$, i,e., the mass of the cocoon is $M_{\rm c} = 0.67 M_\odot$. 
We also take the jet-ejecta interaction time of this case to be $t_{\rm j,0}=7.5\times 10^6 \s = 87 {~\rm d}$, 
and so by equation (\ref{eq:Rc}) with $\beta=0.5$ the initial radius of the gas of the cocoon is $R_{{\rm c},0} = 1.5 \times 10^{14} \cm$. 
The launching time of the jet is given by 
\begin{equation}
t_{{\rm L}} = t_{\rm j,0} - R_{\rm s}(t_{\rm j,0})/v_{\rm L},
\label{eq:t_L}
\end{equation}
where we take the terminal velocity of the jet to be $v_{\rm L} = 10^{10} \cm \s^{-1}$ as in \cite{Gilkis:2015adr}.
The spherical (under our assumptions) expansion velocity of the mini-exploding cocoon relative to its center of mass (that expands with the SN ejecta) is $v_{\rm c} \approx \left(2E_{\rm c}/M_{\rm c}\right)^{1/2} = 950 \km \s^{-1} \ll v_{\rm SN}$. 
We present the light curve of the toy model with the above parameters in Fig. \ref{fig:SN&jet}.
{{{{ The energy $E_{\rm {2j,rad}}$ is the total radiated energy due to the two mini-explosions. We calculate it by integrating the light curve of each mini-explosion (section \ref{sec:jet}) with timescale and peak luminosity from equations (\ref{eq:tjet}) and (\ref{eq:Ljet}), respectively. }}}}

As evident from equation (\ref{eq:Ljet}) the luminosity that result from the jet-ejecta interaction increases with interaction time $t_{\rm j,0}$. The reason is that later interactions occur at larger distance from the center of the SN explosion and in lower density ejecta regions. The large distance implies that the shocked gas in the cocoon requires more time to lose its thermal energy to adiabatic expansion, and the lower ejecta density implies shorter photon diffusion time. Both these effects cause more of the thermal energy of the shocked gas in the cocoon to end up in radiation rather than doing work in adiabatic expansion. 

\begin{figure}[ht!]
	\centering
	\includegraphics[trim=3.5cm 8cm 4cm 8cm ,clip, scale=0.6]{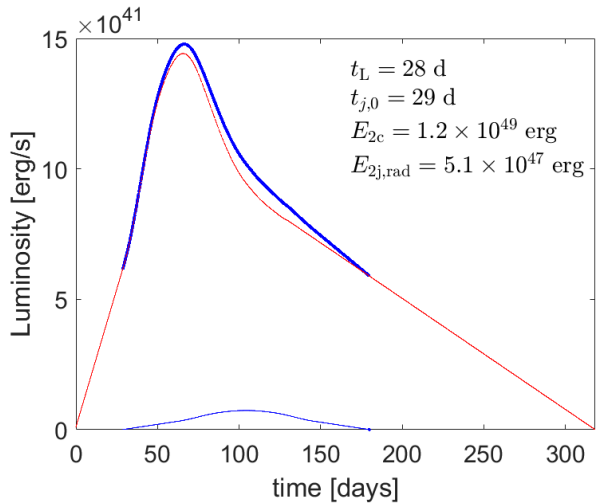}
	\caption{Like Fig. \ref{fig:SN&jet} but for $t_{\rm j,0}=2.5 \times 10^6 \s =29 {~\rm d}$ instead of  $t_{\rm j,0}=7.5\times 10^6 \s = 87 {~\rm d}$.
		The axes are the same as in Fig. \ref{fig:SN&jet}.
	}
	\label{fig:SN&jet2}
\end{figure}

To demonstrate the effect of earlier jet-ejecta interaction we present in Fig. \ref{fig:SN&jet2} a case with jet-ejecta interaction time of $t_{\rm j,0}=2.5 \times 10^6 \s =29 {~\rm day}$,  
keeping all other parameters identical to those of the case we present in Fig. \ref{fig:SN&jet}. 

The total (by the two cocoons) energy radiated in these cases are $E_{\rm 2j,rad} (87 {~\rm day})=1.4 \times 10^{48} \erg$, and $E_{\rm 2j,rad} (29 {~\rm day})=5.1 \times 10^{47} \erg$, respectively. From these numbers and from Figs. \ref{fig:SN&jet} and \ref{fig:SN&jet2} we clearly see that under our assumptions the effect of jets with particular properties increases with later jet-ejecta interaction time. 
     
{{{{{ We further discuss the assumption of a spherical cocoon. In section \ref{sec:jet} we mention the two opposite effects of an elongated cocoon, that of a faster photon escape along the shorter dimension of the cocoon and that of a faster expansion in the shorter dimension of the cocoon. In the case of an elongated cocoon where the faster photon diffusion has a larger effect than the faster adiabatic expansion, the effect would be as if the opacity in equations (\ref{eq:tjet}) and (\ref{eq:Ljet}) is smaller. This would make the jet-induced peak (the mini explosion) brighter and somewhat earlier. An earlier peak means, for a not too early mini-explosion, that it occurs when the background CCSN light curve is brighter. Whether the jet-induced peak will be more prominent depends also on the background light curve. In the case of an elongated cocoon where the faster adiabatic expansion has a larger effect than the faster photon diffusion, the effect would be as if the mini-explosion energy $E_c$ is lower, i.e., $\epsilon_E$ is smaller (eq. \ref{eq:jet2}). From equations   (\ref{eq:tjet}) and (\ref{eq:Ljet}) we find that the jet-induced peak be fainter and occurs later. Despite that the peak is fainter, because it occurs later it might be more prominent even, if the background CCSN light curve decays fast enough. Overall, the effect of a somewhat elongated cocoon is not expected to be huge.   
}}}}}

We can summarize the main result of this section as follows. 
Under our assumptions, jets at late times can have substantial effects on the light curve even when their kinetic energy is a small fraction, about one percent or even less, of the total kinetic energy of the ejecta. The jets can form a large peak after maximum light. At early times the shocked material, the cocoon, has time to adiabatically cool before much of the energy is radiated away as photon diffusion time is long. 

\section{Application to \MakeLowercase{i}PTF14\MakeLowercase{hls}}
\label{sec:iPTF14hlsToy}

\subsection{The enigmatic SN iPTF14hls}
\label{subsec:iPTF14hlsProperties}

One of the properties of the enigmatic transient iPTF14hls (AT 2016bse; Gaia16aog), classified as type II SN, is that there are several late peaks in its light curve \citep{Arcavietal2017Nature, Sollermanetal2019}. Such events might not be extremely rare as \cite{Arcavietal2018ATel} suggest that SN 2018aad (ASASSN-18eo; \citealt{Hosseinzadehetal2018, Nichollsetal2018}) is similar in many aspects to iPTF14hls, {{{{ e.g, a high luminosity, a fast outflow, a slow decay with some rise even, and a faint host galaxy. }}}}

Several theoretical studies propose different scenarios to account for the properties of iPTF14hls, and in particular for its prolonged light curve, but none can fit all properties of iPTF14hls (e.g., \citealt{Sollermanetal2019, Woosley2018}).
These scenarios include a pair instability supernova (e.g., \citealt{Woosley2018, VignaGomezetal2019}), a magnetar, i.e., a rapidly rotating magnetic NS (e.g., \citealt{Arcavietal2017Nature, Dessart2018, Woosley2018}), fallback accretion (e.g., \citealt{Arcavietal2017Nature, Wangetal2018, Liu2019}), an interaction of the ejecta with a circumstellar matter (CSM; \citealt{AndrewsSmith2018, MilisavljeviMargutti2018}), and a common envelope jets supernova scenario (CEJSN; e.g., \citealt{SokerGilkis2018iPTF14hls}). 

Some scenarios for iPTF14hls attribute main roles to jets, in the explosion itself (e.g., \citealt{Chugai2018}), some at late time by fallback accretion onto a NS or onto a black hole (e.g., \citealt{Liu2019}), and some consider jet-powering both in the explosion and at late times, as in the CEJSN scenario \citep{SokerGilkis2018iPTF14hls} and in the scenario of late accretion of hydrogen-rich gas with stochastic angular momentum that power jittering jets \citep{Quataertetal2019}, i.e., the jittering jets explosion mechanism.
Even in the case of magnetar powering, jets most likely play crucial roles in powering the explosion \citep{Soker2016Magnetar, SokerGilkis2017Magnetar}.
\cite{GofmanSoker2019} argue that all iPTF14hls scenarios with jets require a source of angular momentum, and that this source is a stellar companion strongly interacting with the progenitor of iPTF14hls.

Both the accretion fallback \citep{Wangetal2018} and the magnetar model deposit the energy at the center of the SN ejecta. We, on the other hand, deposit the energy further out in the ejecta and in a bipolar morphology rather than a spherical one. In all scenarios the photospheric emission comes from photon diffusion.
Our toy model, therefore, includes two parameters that do not exist in spherically symmetric energy deposition models. These are the relative ejecta mass that is shocked by the jets, $\epsilon_V$ (equation \ref{eq:jet2}), and the half opening angle of each jet $\alpha_j$. 

\cite{Wangetal2018} propose a scenario where intermittent fallback accretion of $\approx 0.2 M_\odot$ explains the late peaks in the light curve of iPTF14hls. Their scenario has an explosion energy of $2.2 \times 10^{51} \erg$ and an ejecta mass of about $21 M_\odot$. 
They could not fit the third peak in the light curve with a fallback model because its short duration. They rather attributed the third peak to magnetic activity of the NS. 
We check below whether the two extra parameters in our toy model might account for the fast third peak. {{{{ Although we study only the third peak that \cite{Wangetal2018} could not account for, in principle it is possible that more peaks come from different jet-launching episodes. }}}}

\subsection{Fitting iPTF14hls peaks with late jets}
\label{subsec:iPTF14hlsFitting}

\begin{figure}[ht!]
	\centering
	\includegraphics[trim=3cm 14.2cm 2cm 0.7cm ,clip, scale=0.58]{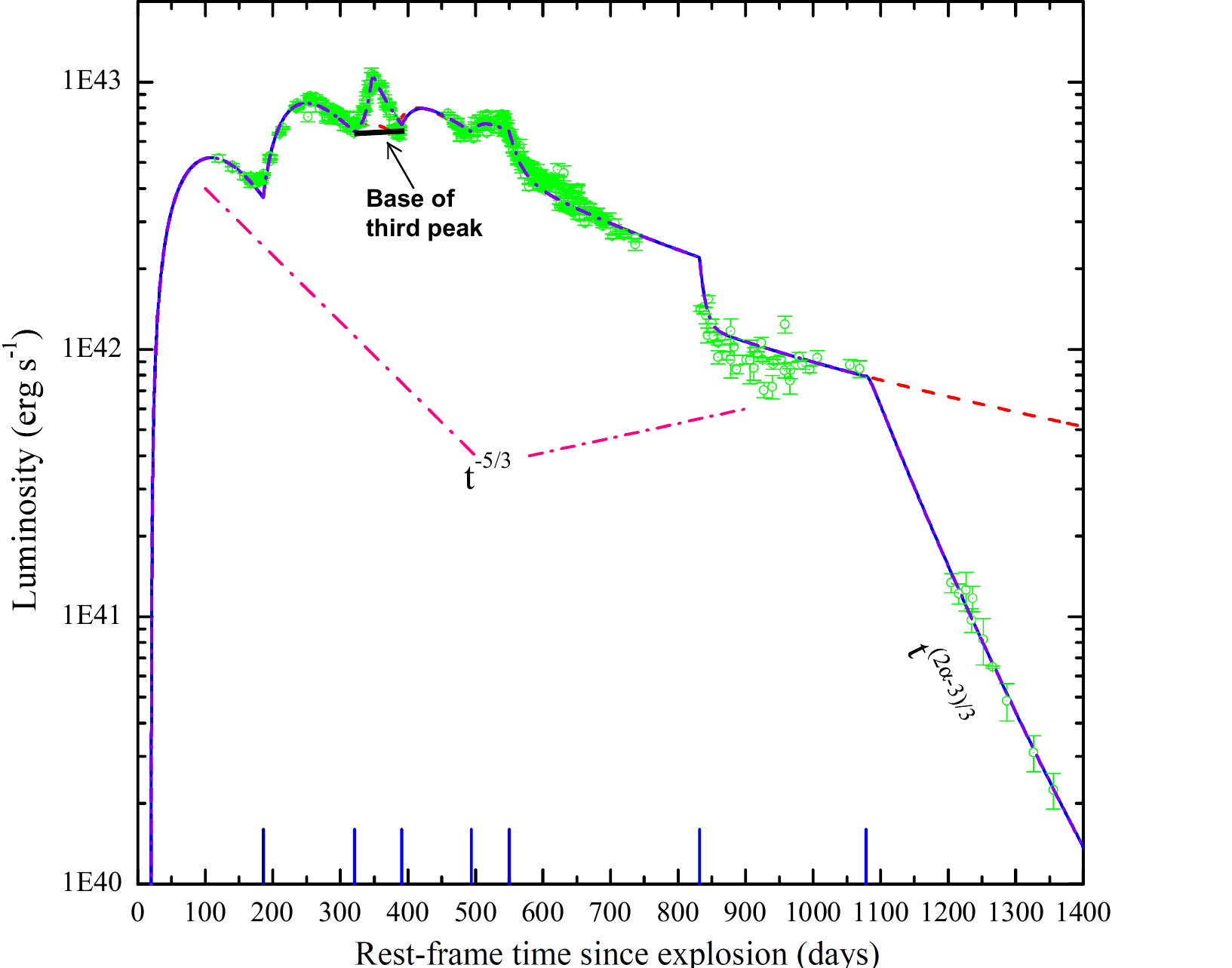}
	\caption{The bolometric light curve of SN iPTF14hls from \cite{Wangetal2018}. 
We mark the approximate base of the third peak that we use, i.e., the contribution from the adjacent peaks and the supernova itself. The total radiated energy of the peak is $\simeq 10^{49} \erg$. The blue lines perpendicular to the horizontal axis mark the beginning time of each fallback episode (i.e., peak) in the modeling of \cite{Wangetal2018}.
	}
	\label{fig:iPTF14hls_Wang}
\end{figure}

We apply our toy model to the strongest peak in the light curve of the enigmatic type II-P SN iPTF14hls, i.e., the third peak that \cite{Wangetal2018} had problems to fit. {{{{ (We note again that our toy model is applicable to other CCSN types, and not only to Type II-P.) }}}}
\cite{Wangetal2018} suggest that this peak is due to a magnetic activity of the newly born NS, in contrast to the other peaks that they attribute to fallback accretion.  
We present the light curve from \cite{Wangetal2018} in Fig. \ref{fig:iPTF14hls_Wang}.

We take the following parameters to our toy model. 
We take the mass and energy of the ejecta, that we need for the background light curve with scaling from equations (\ref{eq:tSN}) and (\ref{eq:LSN}), from \cite{Wangetal2018} to be $M_{\rm SN}= 21 M_{\odot}$ and $E_{\rm SN} = 2.2 \times 10^{51} \erg$, respectively. 
We take the jet-ejecta interaction time of the third peak, which is the starting time of that peak, to be $t_{\rm j,0} =309~{\rm d}$ according to Fig. \ref{fig:iPTF14hls_Wang}. Using our base of the third peak (marked in Fig. \ref{fig:iPTF14hls_Wang}) we calculated its energy, which is the radiated energy of the mini-explosion of our toy model, to $E_{\rm 2j, rad} \simeq 1 \times 10^{49} \erg$. 
We also find from \cite{Wangetal2018} that the time from start to the maximum of the peak, which in our toy model is $t_c$, is 30 days, $t_c=t_{\rm p3}=30 {~\rm d}$.

Using the above values from the light curve, $t_{\rm j,0} =309 ~{\rm d}$, $E_{\rm 2j, rad} \simeq 1 \times 10^{49} \erg$, and $t_c=t_{\rm p3}=30 {~\rm d}$, in equations (\ref{eq:tjet}) and (\ref{eq:Ljet}) with the (symmetric) shape of the peak that we use in our toy model (Section \ref{sec:jet}), we can solve for $\epsilon_E$ and $\epsilon_V$. 
There are two other parameters in the toy model, the distance of the jet-ejecta interaction relative to the SN ejecta, $\beta$, and the half opening angle of the jet $\alpha_j$. 
For these we use the values as in equations (\ref{eq:tjet}) and (\ref{eq:Ljet}), $\beta=0.5$ and $\alpha_j=5^\circ$.

The solution gives {{{{$\epsilon_E=0.0079$}}}} and {{{{$\epsilon_V=0.0133$}}}}. 
From these we find the kinetic energy of the two jets that power the third peak in our toy model to be $E_{\rm 2c}=2\epsilon_E E_{\rm SN}=3.5 \times 10^{49} \erg$, and the half opening angle of the cocoon to be $\alpha_c= 13.2^\circ$  
(Fig. \ref{fig:schecmatic}).

\begin{figure}[ht]
	\centering
	\includegraphics[trim=3.5cm 8cm 4cm 8cm ,clip, scale=0.6]{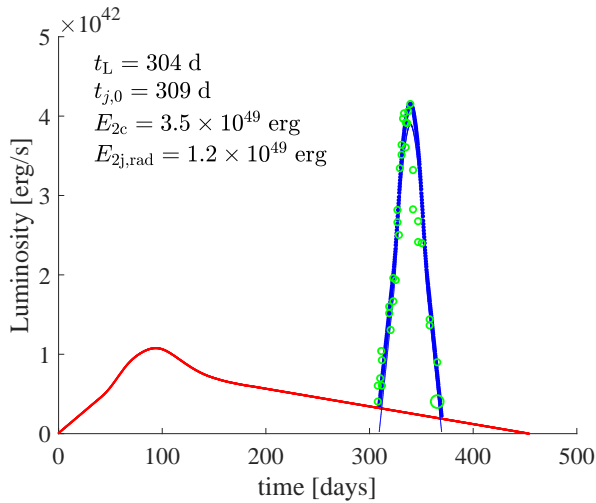}
\caption{The light curve of the SN, {{{{ shape as in Fig. \ref{fig:SNLightCurve} }}}} and the third peak from our toy model where we try to fit the third peak energy and timescale using SN energy and mass from  \cite{Wangetal2018}. The other parameters are in the text (Section \ref{subsec:iPTF14hlsFitting}).  
The red line represents the contribution of the SN itself, the thin blue line represents the 'mini explosion' of our toy model (the peak), and the red + thick blue line represents the total light curve with the peak. 
{{{{The green points are the data of the third peak of iPTF14hls (The Open Supernova Catalog), where the larger green circle on the bottom right of the peak is the general location of several close observations. }}}}
		$t_{{\rm L}}$, $t_{\rm j,0}$, $E_{2{\rm c}}$, and $E_{\rm {2j,rad}}$ are the launching time of the two opposite jets, the 'mini-explosion' time of the jets, the energy that the two jets deposit to the cocoons, and the extra energy radiated by the two cocoons.
	}
	\label{fig:iptf_peak_tot}
\end{figure}

We present in Fig. \ref{fig:iptf_peak_tot} the plot of the SN itself (red line), the peak (thin blue line), the total light curve (red + thick blue line), {{{{and the data of the third peak of iPTF14hls (green circles)}}}}. {{{{ We emphasise that we do not fit the SN light curve, but only the extra energy due to the third peak. }}}}

There are other peaks in the light curve of iPTF14hls (Fig. \ref{fig:iPTF14hls_Wang}), but they are wider, namely active for a longer time, and have more complicated shapes (less symmetric) than the third peak. Our toy model that we build on a mini-explosion does not fit well these wider peaks. For example, they might require jets that active for a long time and with varying intensity. 
We limit ourselves to fit the narrowest and strongest peak, the third peak that \cite{Wangetal2018} had problems to fit with fall back accretion. We here managed to show that we can explained the third peak by fallback accretion if we assume that the accretion process launches jets, rather than a spherically symmetric energy deposition. 
These two jets can explain the brightest peak of iPTF14hls with a  kinetic energy of $E_{\rm 2c}=3.5 \times 10^{49} \erg$ that is only about $1.6 \%$ of the supernova energy. 
     
{{{{ We can crudely estimate the required fallback mass to explain the third peak as follows. We assume that the two jets carry a fraction of $\approx 0.1$ of the accreted mass in the fallback episode, and that their terminal velocity is about the escape velocity from an accretion disk around the NS, namely $v_j \simeq 10^5 \km \s^{-1}$. For these parameters the accreted mass in the third-peak episode is $M_{\rm acc,3} \simeq 0.0035M_\odot$. This is compatible with the total fallback mass of $\approx 0.2 M_\odot$ that \cite{Wangetal2018} estimate in their model for all peaks. This is because we think that the interaction where jets penetrate into the ejecta, something impossible in a spherical mode, is more efficient in converting kinetic energy to radiation. }}}}

\section{Summary}
\label{sec:summary}

We built a toy model to explain peaks in the decline phase of the light curves of CCSNe. In this toy model we assume that the central object launches two short-lived opposite jets that catch up with the CCSN ejecta. The collision shocks the jets' material and the ejecta to form hot bubbles, the cocoons (Fig. \ref{fig:schecmatic}). We refer to this interaction as `mini explosion', and calculate its influence on the light curve by further assuming a spherically symmetric cocoon. 
{{{{ In sections \ref{sec:jet} and \ref{sec:Peaks} we discussed that the assumption of spherically symmetric cocoon is not a strong assumption. Namely, an elongated cocoon will not affect much our results. }}}}
We use results from \cite{Kasen09} to describe the timescale and energy output of both the CCSN itself (equations \ref{eq:tSN}, \ref{eq:LSN}) and of the `mini explosion' (equations \ref{eq:tjet}, \ref{eq:Ljet}). {{{{ We listed the assumptions that go into these equations and others in Table \ref{Table:Parameters}. }}}} For the shape of the light curve of the CCSN itself (without late jets) we constructed the light curve that we presented in (Fig. \ref{fig:SNLightCurve}). {{{{ This artificial light curve includes all contributions beside the late jets, namely, the explosion thermal energy of the ejecta, radioactive decay, and recombination of the ejecta. }}}} We assumed that the {{{{ rise side of the peak (extra radiation) that the jets form has the shape as the peak of the SN itself (but different timescale and energy as listed in Table \ref{Table:Parameters}), and that the decay side is symmetric to the rise side. }}}}

At late times the jets form a much stronger peak than at early times because then the interaction of the jets with the SN ejecta occurs at a larger distance from the center of the SN explosion, and at a lower density ejecta region. As a result of that the shocked gas in the cocoon takes more time to lose its thermal energy to adiabatic expansion and the photon diffusion time is shorter due to the lower ejecta density. Both of these act to channel more of the thermal energy of the shocked gas to radiation. 
We present the light curve resulting from a late jet-ejecta interaction in our toy model in Fig. \ref{fig:SN&jet}, and an early one that has a much less influence on the light curve in Fig. \ref{fig:SN&jet2}.

{{{{ A series of frequent jet launching episodes might produce a plateau in the light curve by causing a series of `mini-explosions' one after the other. We expect the plateau not to be smooth, but rather to have a wavy structure. }}}}

{{{{ We can apply our model to all types of CCSNe that show late peaks. }}}} Here we applied it to the enigmatic SN iPTF14hls that has several late peaks in its light curve that are yet to be fully explained.
We apply our toy model to the third peak of iPTF14hls (we mark this peak on Fig. \ref{fig:iPTF14hls_Wang}), which is the strongest and narrowest peak in this light curve.
We take the CCSN ejecta mass and energy from \cite{Wangetal2018}, and use equations (\ref{eq:tjet}) and (\ref{eq:Ljet}) to solve for the energy of the jets and the mass of the ejecta that the jets shock. Namely, we solve for the two parameters $\epsilon_E=0.0079$ and $\epsilon_V=0.0133$ that are defined in equation (\ref{eq:jet2}).
From these we find the kinetic energy of the two jets that power the third peak in our toy model to be $E_{\rm 2c}=2\epsilon_E E_{\rm SN}=3.5 \times 10^{49} \erg$, which is only about $1.6 \%$ percent of the total kinetic energy of iPTF14hls $E_{\rm SN}= 2.2 \times 10^{51} \erg$ (from \citealt{Wangetal2018}).
The jets shock a mass of $M_{\rm 2c}=0.56 M_\odot$ out of total ejecta mass of  $M_{\rm SN}=21 M_\odot$ (from \citealt{Wangetal2018}). 
{{{{ We present our fitting to the third peak in Fig. \ref{fig:iptf_peak_tot}.
We emphasize that we do not try at all to fit the light curve of SN iPTF14hls, and therefore the red line in Fig. \ref{fig:iptf_peak_tot} has no real meaning, but it rather serves only as a background light curve. }}}}
We note that the jets are launched by fallback accretion, and so we actually can explain the third peak with fallback accretion that launches jets. 

Our general conclusion is that jets can explain peaks that spherically deposition of energy in the SN ejecta cannot explain. At late times the energy of the jets can be only a small fraction of the total SN energy, as we showed for the third peak of iPTF14hls where the jets carry only about one to two percent of the CCSN energy. 

\section*{Acknowledgments}
{{{{ We thank an anonymous referee for very detail comments that substantially improved the presentation of our results. }}}}  This research was supported by a grant from the Israel Science Foundation.

\label{lastpage}
\end{document}